\documentclass[11pt]{article}
\usepackage[noadjust]{cite}
\usepackage[cmex10]{amsmath}
\usepackage[pdftex]{graphicx}
\usepackage{color}

\renewcommand\footnotemark{}

\begin{document}
\title{\vspace{-2cm}\bf{Smart Detector Cell: A Scalable All-Spin Circuit for Low Power Non-Boolean Pattern Recognition}}   % type title between braces

\author{Hamidreza Aghasi${}^\dagger$,
Rouhollah Mousavi Iraei${}^\star$, 
Azad Naeemi${}^\star$ and
Ehsan Afshari${}^\dagger$ 
\footnote{${}^\dagger$School of Electrical and Computer Engineering,
Cornell University, Ithaca, NY 14850 }\footnote {${}^\star$School of Electrical and Computer Engineering,
Georgia Institute of Technology, Atlanta, GA 30332}
}

   % type author(s) between braces
\date{}    % type date between braces

\maketitle

\begin{abstract}


% for Transactions on Magnetics papers, we must declare the abstract and
% index terms PRIOR to the title within the \IEEEtitleabstractindextext
% IEEEtran command as these need to go into the title area created by
%\IEEEtitleabstractindextext{%\begin{abstract}
\textit{Abstract}---We present a new circuit for non-Boolean recognition of binary images. Employing all-spin logic (ASL) devices, we design logic comparators and non-Boolean decision blocks for compact and efficient computation. By manipulation of fan-in number in different stages of the circuit, the structure can be extended for larger training sets or larger images. Operating based on the mainly similarity idea, the system is capable of constructing a mean image and compare it with a separate input image within a short decision time. Taking advantage of the non-volatility of ASL devices, the proposed circuit is capable of hybrid memory/logic operation. Compared with existing CMOS pattern recognition circuits, this work achieves a smaller footprint, lower power consumption, faster decision time and a lower operational voltage.
\end{abstract}


%\IEEEdisplaynontitleabstractindextext

% \fi

\section{Introduction}
Pattern recognition and in particular, image recognition techniques have been widely studied in machine learning and image processing \cite{pattern1,pattern2, pattern3}. Hardware demonstration of computation units for pattern recognition; however, has consistently been a challenging problem in terms of chip size, power consumption, computation complexity and decision speed.

Among different solid state technologies, CMOS provides the chance of low cost, highly-integrated low power implementation for pattern recognition \cite{cmos2, simple, nuero} and processing \cite{cmos3, Liu} systems. For boolean logic systems, CMOS gates exhibit processing speeds up to  a few GHz and can be designed to have a low static power. However, the dynamic power consumption of a large system with a GHz clock frequency can still limit the scalability. Fan-in and fan-out considerations for CMOS devices also impact the speed, power consumption and the size of devices. Besides boolean systems, some novel non-Boolean techniques have been developed to overcome these issues. In non-Boolean systems, logic gates will no longer be the key block and analog/mixed signal circuits are used. In \cite{simple}, the authors propose a technique for non-Boolean training and detection of image pixels using a network of coupled oscillators. This structure has the capability to detect any scaled or rotated version of a desired image. On the other hand, this method suffers from high computational complexity, large area and high power consumption which limit the application for large image arrays. To this, we should also add the long convergence time. Other proposed CMOS systems have demonstrated artificial neural networks (ANN) by designing circuits emulating neurons and synapses \cite{cmos2, nuero}. In these systems, the larger computation burden, leaves the search open for new solutions.

To overcome the limitations of CMOS devices, other technologies are being investigated for pattern recognition applications. Spintronic devices, in particular, have received a lot of attention recently because of some unique properties, e.g., low voltage operation and non-volatility. In \cite{Fabrication}, a non-volatile logic-in-memory full adder is fabricated using the magnetic tunnel junctions (MTJ's). The proposed architecture is compared with an 0.18$\mu$m CMOS process counterpart and exhibits major advantages. The dynamic power consumption compared to a conventional CMOS circuit is 23$\%$ reduced due to reduction in the number of paths from $V_{DD}$ to GND. On the other hand the static power consumption is eliminated due to the non-volatility and the chip area is also reduced. As stated in \cite{Dimitri}, the ASL devices are also non-volatile and the conputational state is preserved when the power to the circuit is turned off. In \cite{Sharad}, a spin-based artificial neural network (ANN) is proposed using lateral spin valves to achieve a low power consumption and a low operatoinal voltage. In \cite{LAST}, spin swithces to develop compact neurons and synapses are proposed. In \cite{datta}, all-spin logic (ASL) and  charge-spin logic (CSL) devices are shown to be capable of Boolean and non-Boolean operations which make them an attractive choice to build some fundamental blocks such as ring oscillators. Recently, the design of ASL gates with graphene channels have been proposed \cite{graphene}. Due to the unique features of graphene in terms of the spin transport, the design of Boolean and non-Boolean computation units with these new devices can be investigated as a future direction. Moreover, majority gate operation of ASL devices has been previously introduced in some Boolean logic systems \cite{Behin-Ain, Augustine}. This unique feature of these devices can overcome fan-in and fan-out limitations of large integrated systems. Besides, the inverting and non-inverting operation modes of ASL devices can be the key to design many logic circuits e.g., full adder circuits and multipliers \cite{adder}. The time domain transient behavior of magnetization in these devices can also provide another degree of freedom to demonstrate non-Boolean operations as it will be discussed later in this paper. These features, enable us to design an all-spin logic non-Boolean compact structure with low power consumption and low computational complexity.

In this paper we propose a novel pattern recognition circuit that takes advantage of most novel features of spintronic devices such as non-volatility, efficient implementation of majority gates and XOR functions and the ability to distinguish strong and weak majorities. Non-volatility of the devices enables storing large sets of training images within the logic with no standby power dissipation. This feature also enables ‘instant-on’ operation and saves on energy and delay penalties imposed by loading training images form a main memory.

The rest of this paper is organized as follows. Section II describes the operation of ASL devices. The proposed approach and the basic of computation are given in Section III. In Section IV, the proposed architecture and a comprehensive discussion on design considerations are presented.  Simulation results and summary are shown in Sections V and VI, respectively.

\section{All-Spin Logic Devices}
Spin of electron is introduced as a new state variable in spintronic devices to process and store information. This new alternative to charge-based systems, provides the possibility of achieving an ultra-low voltage operation and easier demonstration of digital systems coming from the bistable nature of spin \cite{Ultra}.

In all-spin logic devices, input and output data are represented by the magnetization of two ferromagnets \cite{azad} which are communicating through a spin-coherent metallic channel. The physical view of these devices is shown in Fig 1(a). As shown in the Fig. 1(b-d) and discussed in \cite{impact}, the applied voltage on the input ferromagnet, creates a flow of electrons which moves them from the supply voltage to the ground. This flow of electrons, becomes spin-polarized when passing through the input ferromagnet. Since the concentration of the spin-polarized electrons are different at the input side and the output side of the channel, the electrons diffuse to the output side. The accumulated spin-polarized electrons under the output ferromagnet, can switch the magnetization orientation of the magnet by applying a torque based on spin-transfer torque effect.
\begin{figure}[h]
\centering
\begin{tabular}{c}
\includegraphics[width=.5\textwidth]{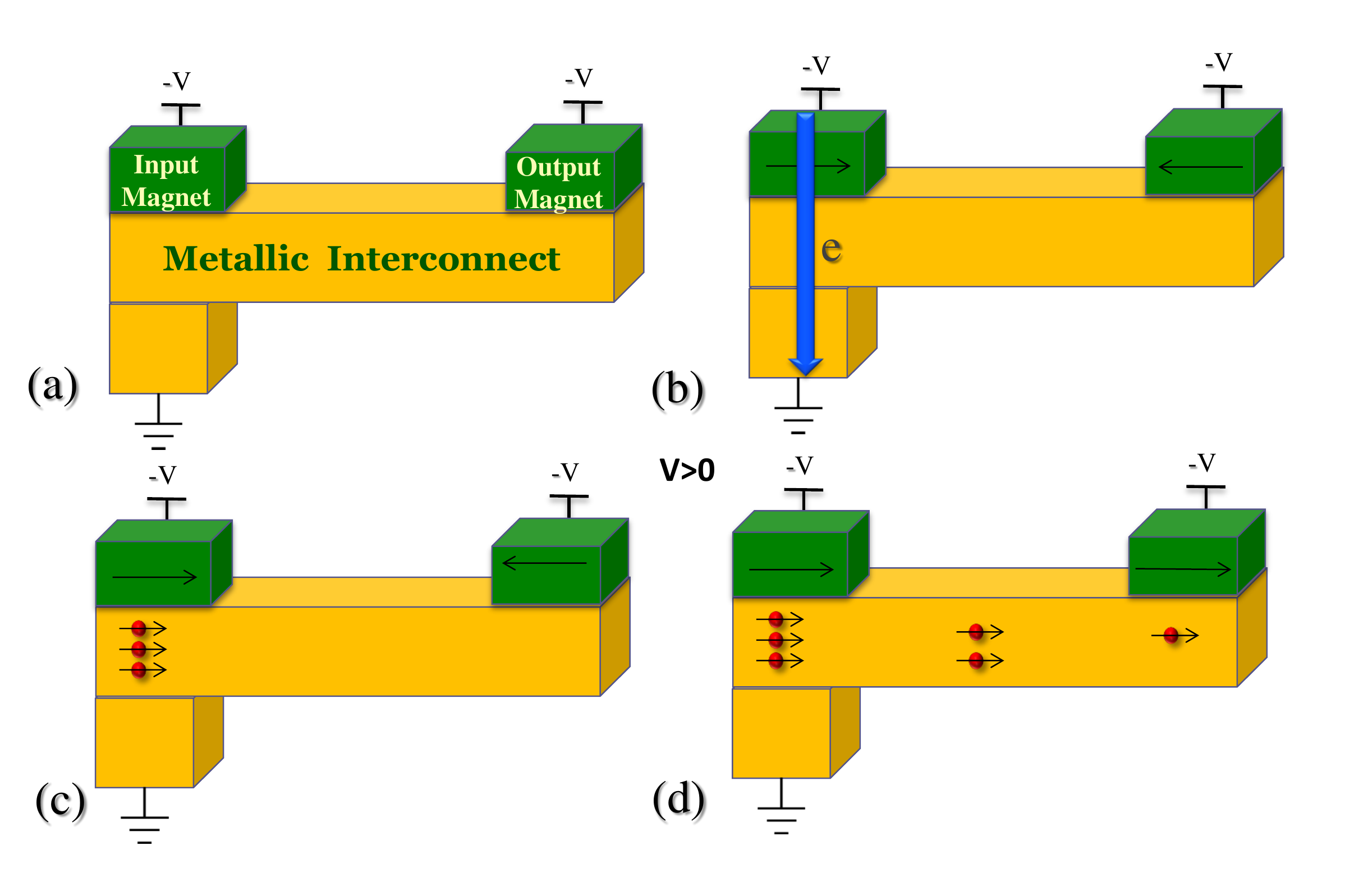}\\
\includegraphics[width=.4\textwidth]{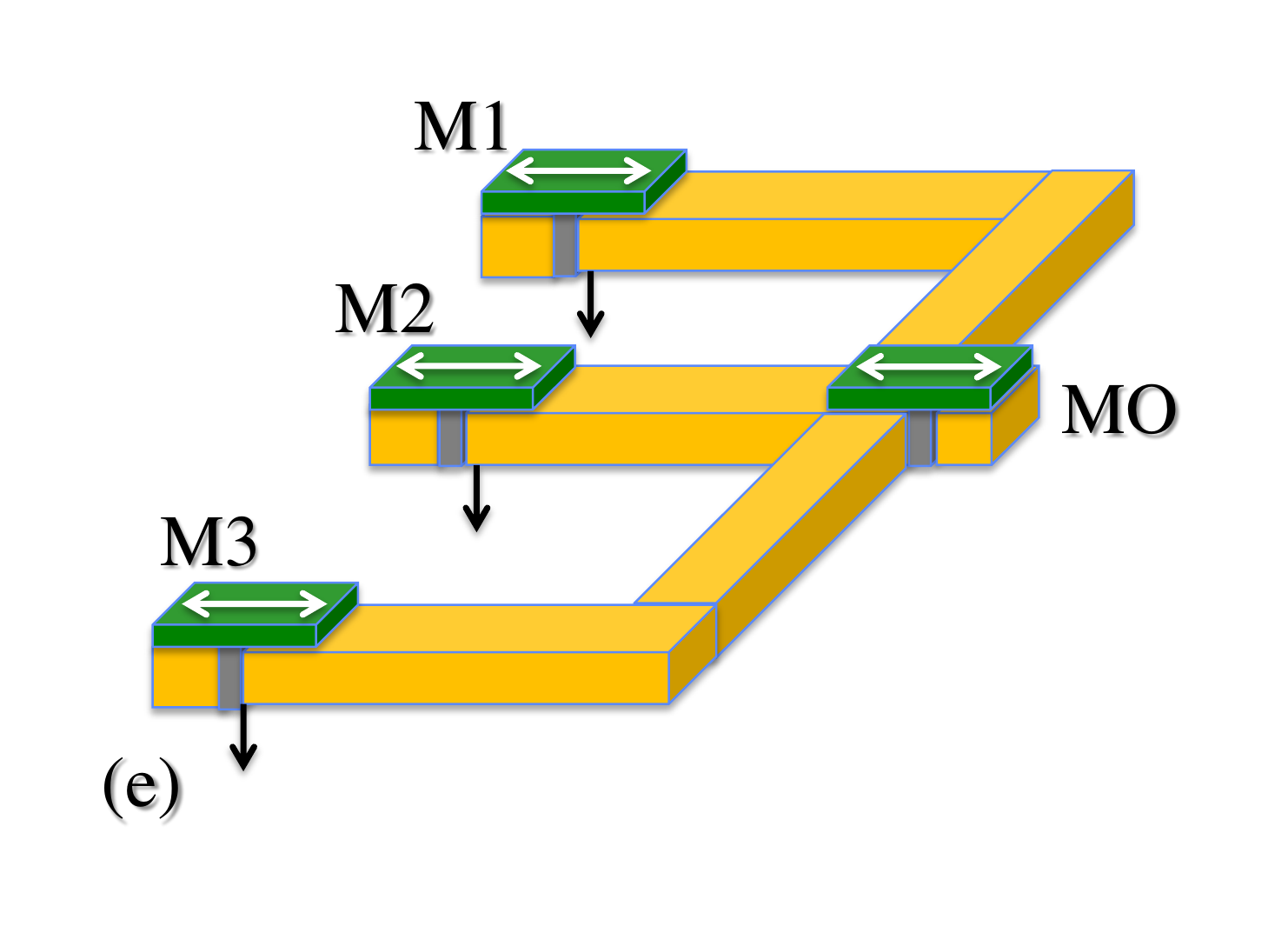}
\end{tabular}
\caption{(a) Configuration of single ASL device (b) Applied voltage on the magnet, creates an electric field and enforces electron movements. (c) Spin-polarized electrons at the input side exhibit a higher density compared to the output side. (d) The diffusion of spin-polarized electrons towards the output magnet, changes the output magnetization direction. (e) An ASL Majority Gate with 3 inputs \cite{azad}. The 3 input magnets, M1, M2 and M3 are connected to the output magnet MO using 3 metallic interconnects.}
\end{figure}

As shown in \cite{Behin-Ain}, these devices can be concatenable, exhibit nonlinear characteristics and support all Boolean operations. In all-spin logic operation, by using the direct spin signal, the nanomagnet can be switched and this signal can be transferred to the next stage. By storing the information in the spin magnetization of magnets, the input and output magnets can effectively be considered as digital capacitors linked by a spin-coherent channel. The sign and magnitude of control voltages applied on the magnets, determines the polarity of majority spin electrons and the device speed, respectively. Any change of magnetization in the bistable input magnet can exert a spin current through the channel and this current can determine the spin magnetization of the output magnet \cite{azad}. The channel between the two magnets can be either a metal or a semiconductor \cite{metal}. In our modeling and simulations we assume a copper interconnect.\\

The models utilized in this work are based on \cite{azad} where the different physical effects are captured. The accurate parameters of channel, magnet and interface that determine different performance characteristics, e.g., the spin injection, detection and transport efficiency are taken into account. The main important size effect parameters for the purpose of this work are the side wall specularity, the grain boundary reflectivity and the average grain size \cite{impact}. The average grain size is assumed to be equal to the width of thickness of the metals \cite{impact}. The complete list of parameters is in  Appendix B.
\subsection{Majority Gate Operation}\label{sec:major}
 As mentioned earlier, the ASL device supports a majority operation as shown in Fig 1(e). This feature is achieved because the net spin current to the output magnet can be determined by the sum of all input spin currents from all the input devices. In principle, this system can be designed for large number of inputs. As a trade off, by increasing the number of input devices in a majority gate, the uncorrelated thermal noise of these devices add up and impact the transient magnetization of output magnet, thus we need to make sure that in the design we have the proper fan-in. As it will be discussed later, if we only want to monitor the final steady value of the output magnetization,  we can keep increasing the number of input devices  as far as the output magnetization is predictable. Based on the device properties, this phenomenon sets a practical limit on the number of input devices to a majority gate. On the other hand, if we care about the transient behavior of the output magnetization, fewer inputs should be connected to the output magnet to avoid the noise accumulation of input devices. In our simulations, for 3 and 5 input cases, the transient output magnetization is less impacted by the thermal noise, compared to higher fan-in numbers. We have to clarify that the steady state value of the majority gate depends on the sign of applied voltage on the magnets. In case of having a negative voltage applied on the magnets, the magnetization orientation value will finally be the exact majority of the input magnetizations. However, if the applied voltage is positive, the steady state value of the output magnet will be the complementary majority of the input magnetizations. \\

The interesting phenomenon in ASL majority gates rises from the dependency of the transient behavior of the output magnetization on the number of similar input magnetizations. This effect can be validated by the fact that the transferred spin torque increases when there are more magnets with magnetization in the same direction. Fig. 2, shows the different scenarios of transient output magnetization in majority gates with 3 and 5 inputs. As shown in Fig. 2(a), in a majority gate, with 5 inputs, the switching of output magnetization becomes faster when there are more inputs with similar magnetization directions. As the number of magnets with similar magnetization decreases, the switching happens slower and the effect of thermal noise is sensed more. In Fig. 2(b), the switching transition for two majority gates with 3 inputs and 5 inputs are compared. By considering the fact that the thermal noise accumulation in the gate with 3 inputs is less compared to the gate with 5 inputs, in the case of having equal net spin currents to both gates, the gate with 3 inputs, exhibits more deterministic transition.
\begin{figure}[!t]
\centering
\includegraphics[width=.75\textwidth]{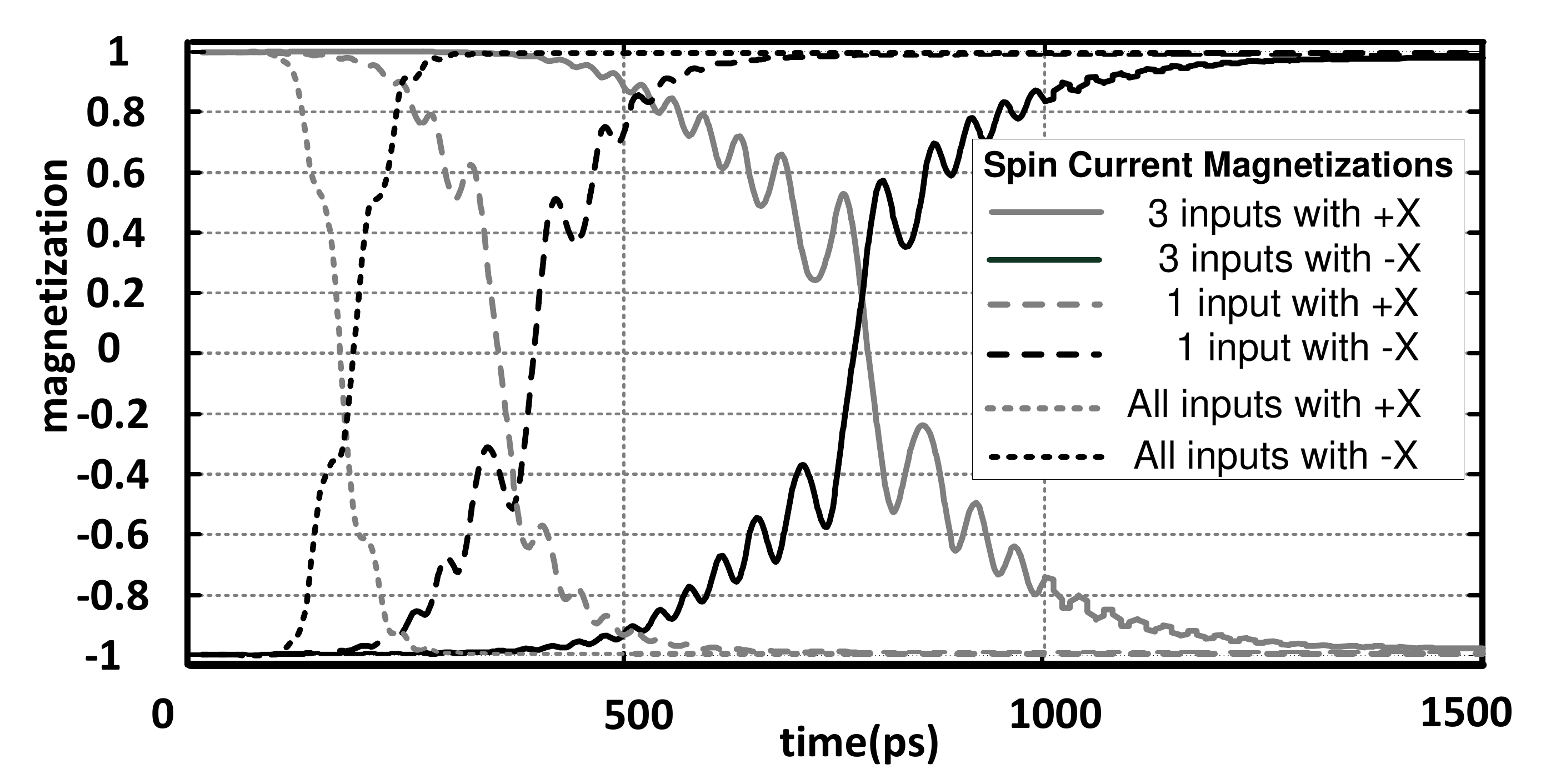}
\includegraphics[width=.75\textwidth]{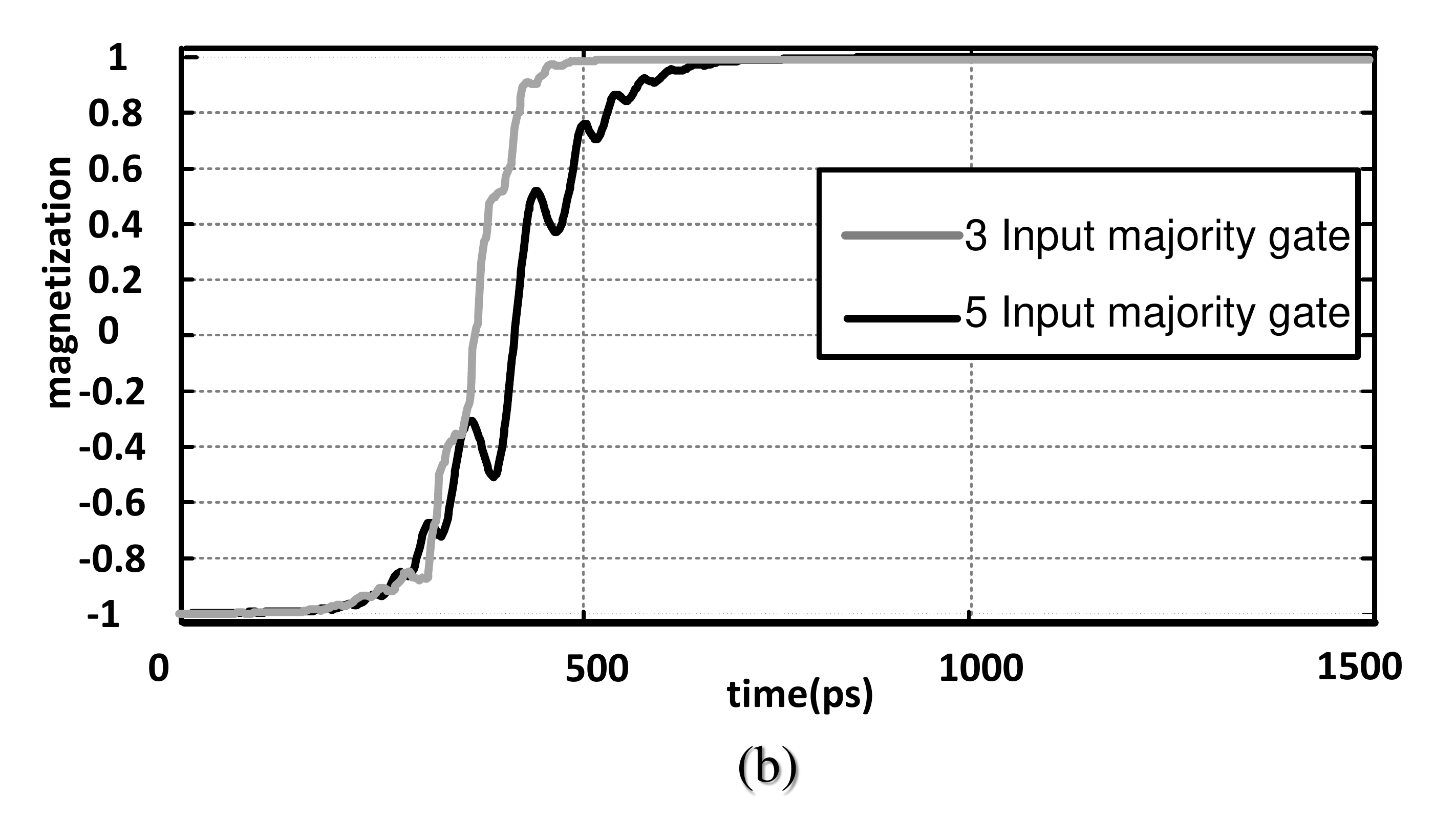}
\caption{(a) Switching transient response for different scenarios of input magnetization in a majority gate with 5 inputs. (b) Switching transition comparison of majority gates with 3 and 5 inputs. In this comparison, the input magnetization of magnets to the 3 input gate are all similar. For the gate with 5 inputs, 4 inputs have similar magnetization and the net spin current is equal to the other gate. The applied voltage on the magntes in these simulations is $-5$ mV.}
\end{figure}
\subsection{Switching Delay Variation}
The switching time of a ferromagnet is calculated in \cite{delay} using the small cone-angle approximation
\begin{equation}
\tau_{s,\omega}=\tau_0\frac{ln(\pi/\theta_0)}{\chi-1},
\end{equation}
where $\tau_0$ is a fitting parameter, $\theta_0$ is the initial angle of the magnet, and $\chi$ is the ratio of the magnitude of injected spin current to the critical spin current required for the switching of the magnetization of the output ferromagent. Based on this equation, if the value of injected spin current increases, the switching delay decreases. However, as shown in \cite{azad}, the channel in this device can be approximated as an RC network; hence, the injected spin current and the value of supply voltage are directly correlated. Therefore, the switching delay is inversely proportional to the value of supply voltage. This result is shown in Fig. 3. The device parameters used for these simulations are shown in 
Appendix B.
\begin{figure}[!h]
\centering
\includegraphics[width=.7\textwidth]{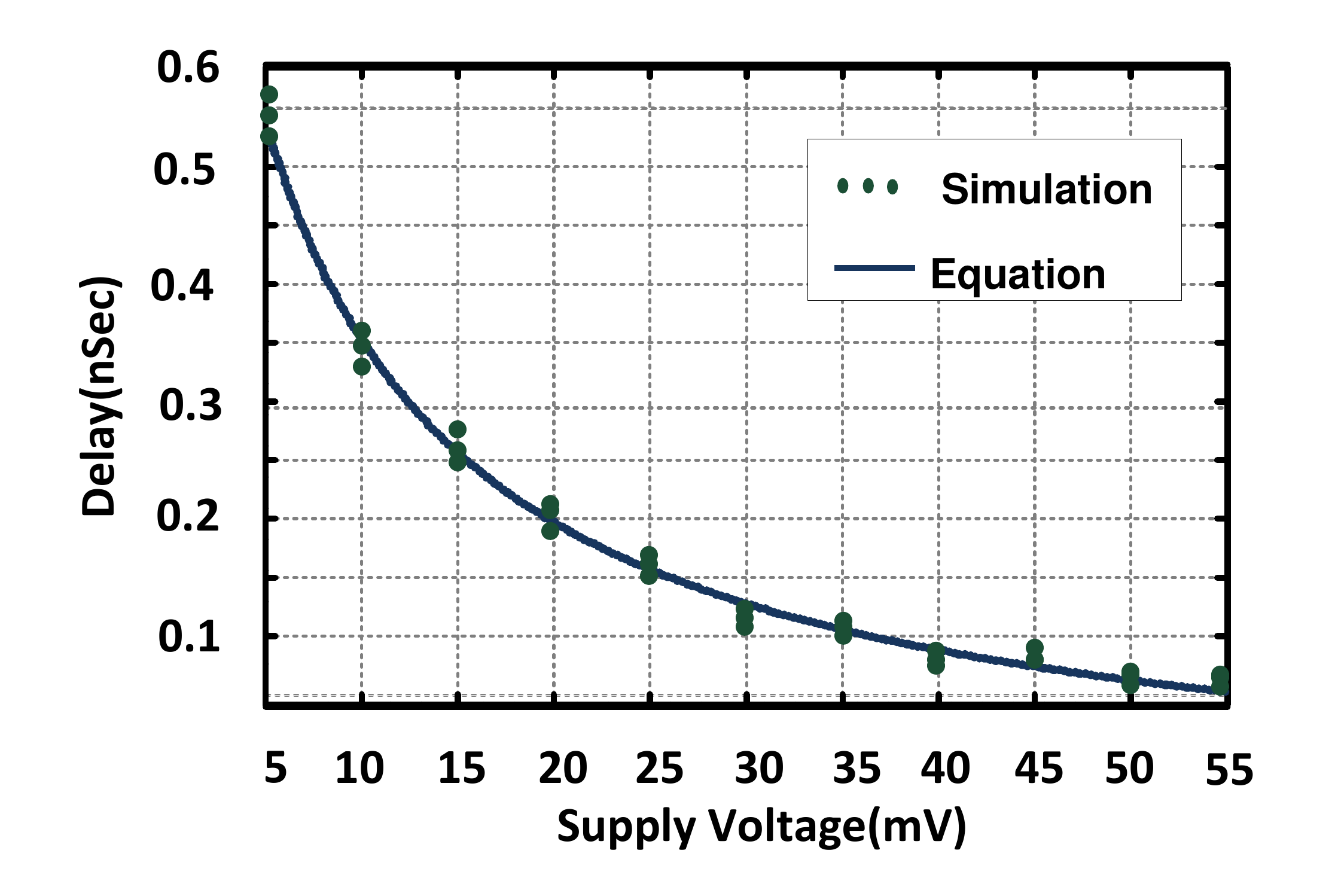}
\caption{Switching delay variation versus the supply voltage. Each point is simulated 3 times to verify the results.\cite{adder}}
\end{figure}
\subsection{ Impact of Thermal Noise}
The thermal motion of electrons inside the ferromagents, is the main cause of thermal noise. The amount of noise is correlated to the temperature of magnets and can directly affect the steady-state precession angel $\theta_0$. Based on the derivations in \cite{noise}
\begin{equation}
 <{\theta_0^2}>=\frac{K_bT}{E_b}.
\end{equation}
In this equation, $E_b$ is the barrier energy and $T$ represents the temperature. This thermal effect acts as the main source of noise which can impact the output magnetization. In our simulations, $\theta_0$ can differ 5$\%$ to 10$\%$  from the analytic solution based on different parameters.
\section{Pattern Recognition Scheme}
Similar to any recognition system, in this work we consider two major phases for the operation. The first phase is the ``learning'' phase where the desired pattern is stored in memory. In ``detection'' phase, the circuit identifies the similarity of an input data and the stored pattern with respect to the decision making criteria. In the learning phase, the circuit can receive a single image or a training set. The training set includes multiple training images from different users.

 In this section, we propose a new technique using all-spin logic devices and establish a fully spin-based operation. By illustrating several examples, we verify the performance for various image sizes.
 
\subsection{Mainly Similar Images}
We first provide the mathematical definition of mainly similarity and then show how this can help the training of the circuit. In our simulations, all the images are binary-valued matrices with 0 and 1 representing white and black pixels, respectively. In our circuit, we assume that binary ``0'' logic corresponds to magnetization orientation in $-X$ direction and binary ``1'' logic corresponds to magnetization orientation in $+X$ direction.

For a given pair of binary vectors $x$ and $y$ with equal length $L$, the \emph{Hamming distance} \cite{Robinson} is defined as 
\[
d(x,y)=\sum\limits_{i=1}^L 1-\delta_{x_iy_i},
\]
where $x_i$ and $y_i$ denote the $i^{th}$ components of $x$ and $y$ respectively and $\delta$ is the kronecker delta. Subsequently, we can exploit this quantity as a measure of similarity between two images. \\

\textbf{Definition 1} \emph{Two binary images $B$ and $B'$ $\subset \{0,1\}^{m\times n}$ are called mainly similar if the majority of pixels across every two rows are identical. More specifically,}
\begin{equation}
\forall k\in\{1,\cdots, m\}: \qquad  d(B_{k,:},B_{k,:}')< \lfloor{\frac{n}{2}}\rfloor, 
\end{equation} 
\emph{where $B_{k,:}$ denotes the $k^{th}$ row of $B$ and $\lfloor a \rfloor$ represents the floor operation on $a$ (i.e., the largest integer not greater than $a$). }

By this comparison, we ensure that the two images have almost similar pixels along the corresponding rows. For the purpose of this paper, we considered the comparison along the rows, although a column-wise comparison could be established with no loss of generality. As illustrated in Fig. 4, being  mainly similar along the rows, does not imply being similar along the columns. 
\begin{figure}[!h]
\centering
\includegraphics[width=.65\textwidth]{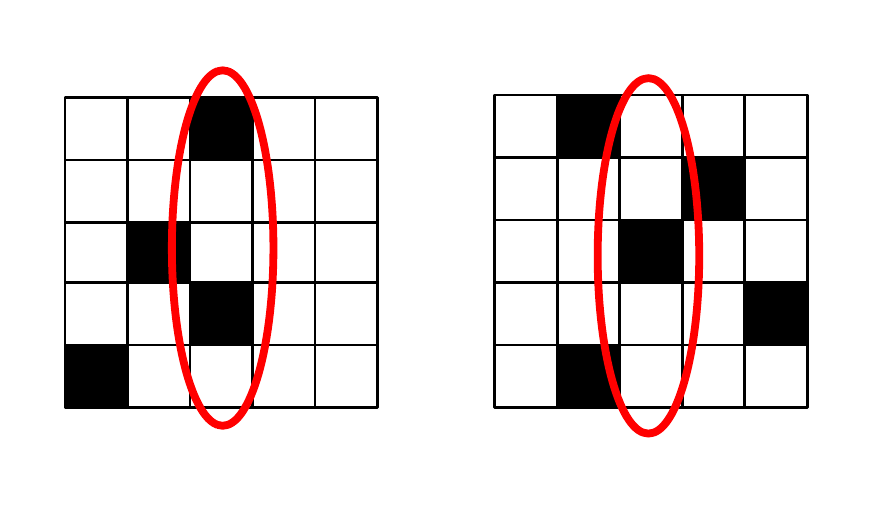}
\caption{The two images are mainly similar (along the rows), however, the Hamming distance between the third columns is 3 which does not imply a similarity along the columns}
\end{figure}
\\
%\textbf{Definition 2}  A set of binary images B=$\{B_1,B_2,...B_P\} \subset \{0,1\}^{m\times n}$ are called majority similar with respect to a binary Pattern Matrix $Q\subset R^{m\times n}$ if the members are majority similar and for any member $B_i$
%\begin{equation}
%d_k(i,Q) \leq \lfloor{\frac{n-||Q_k||_1}{2}}\rfloor+1
%\end{equation}
%Where $||Q_k||_1$ represents the 1-norm of the $k^{th}$ row of Pattern Matrix Q, k is the corresponding row number and $d_k(i,Q)$ is the hamming distance of kth row of matrix $B_i$ and Pattern Matrix Q. By this definition, we make sure that for a row of pattern image which has more black pixels, the distance of images $B_i$ should be less. As the number of black pixels in a row decreases, that represents a lower priority row; hence, the distance can be more. Images shown in Fig 5(b) are majority similar with respect to Pattern Q in Fig 5(a).
%Here I should change it to a and b and use tabular
%\begin{figure}[!t]
%\centering
%{\includegraphics[width=.2\textwidth]{similar2a.pdf}
%{\includegraphics[width=.2\textwidth]{similar2b.pdf}
%\label{fig_second_case}}
%\caption{(a) Desired pattern and (b) a set of majority similar images with respect to this pattern }
%\label{fig_sim}
%\end{figure}
\subsection{Majority Training and Decision Making}
In the learning phase, we train the circuit by providing a number of mainly similar images. In reality, these images could be different representations of a target image (say a character or a certain binary pattern). We build up a representative of the given similar images by constructing a so-called \textit{mean image}.\\

\textbf{Definition 2} \emph{For a set of $P$ binary images $B_1,B_2,\cdots,B_P$ $\subset \{0,1\}^{m\times n}$, the corresponding \textit{mean image} denoted as $\bar B$ is a binary image with entries}
\begin{equation}
\bar B(i,j)=nint(\frac{1}{P}\sum\limits_{k=1}^P B_k(i,j)).
\end{equation}
In this equation, $nint$ denotes the nearest integer function. In our circuit, the mean image represents the desired pattern by the users and is utilized as a reference. Since this matrix is constructed using all-spin majority gates, the number of training images, $P$, is considered to be odd and upper bounded by the maximum number of inputs to a majority gate as discussed in subsection \ref{sec:major}. 

After the training data is stored and the mean image is constructed, we make a row-wise comparison between the input and the mean image. As we will see in the next section, depending on the initial value of output magnetization, the non-Boolean row decision maker can return the total count of matches or mismatches between the compared rows of input image and the mean image.
%In this scheme, the training is considered correct, if for any pixel $(i,j)$ 
%\begin{equation}
%|P\times Q(i,j)-\sum\limits_{k=1}^P B_k(i,j)|\leq \lfloor\frac{P}{2}\rfloor.
%\end{equation}
%Here P is the number of training images. This constraint ensures that for any pixel of interest, the total information from the training set would not lead to a decision error. Accordingly, the probability of false training for an arbitrary pixel is
%\begin{equation}
%\frac{\sum\limits_{i=0}^{P-\lfloor\frac{P}{2}\rfloor}{P \choose \lfloor\frac{P}{2}\rfloor+i}}{2^{nP}}.
%\end{equation}
%The constraint of majority similarity ensures that for other pixels, the conditional probability becomes smaller; hence, the probability of correct detection increases.
%\subsection{Decision Making}\label{sec:Pdef}
% The probability of having $q$ mismatches between the corresponding rows of the input image and the mean image, dented as $\gamma_q$ is calculated to be
%\begin{equation}
%\gamma_q=\frac{{n \choose Q}}{2^{(P+1)n+Q}}.
%\end{equation}
%As it is verified by this equation, the more the number of images in the training set, the less the probability of mismatches will be. As mentioned previously, the maximum value of $P$ is determined by the practical limit coming from the sensitivity of majority gates to thermal noise of input magnets.}
\section{Proposed Structure and Design Considerations}
Based on the pattern recognition scheme shown in Section III, we study two different implementations of the circuit. By comparing the performances of the two different versions of the \textit{single pixel comparator} unit, we choose the one with more capabilities, at the expense of slightly more power consumption and occupied area. In the single pixel comparator, the circuit receives the training pixels from $P$ different users and the mean image is constructed, subsequently. The value of the mean pixel is then compared with the corresponding value in the input image and the steady state magnetization of \textit{Pixel} magnet stores this information.  The two versions of this unit both operate based on the idea of training the circuit with a set of mainly similar images and comparison of the single pixels from the input image with their correspondence in the mean image. With respect to the required operations, the single pixel comparator, needs a memory to store the training data, a logic comparator and a circuit to construct the mean pixel. As previously mentioned, the mean pixel can be constructed by an all-spin majority gate; however, for the memory and the comparator, we will propose a new circuit in the following subsection.\\
\subsection{Memory+Logic Comparator}
1-bit full adder structures with a total number of 5 nanomagnets have been designed in \cite{adder} and \cite{Roy}. By proper setting of the circuit in \cite{adder}, we use it as an area and power efficinet comparator (XNOR) block as shown in Fig 5.
\begin{figure}[!h]
\centering
\includegraphics[width=.75\textwidth]{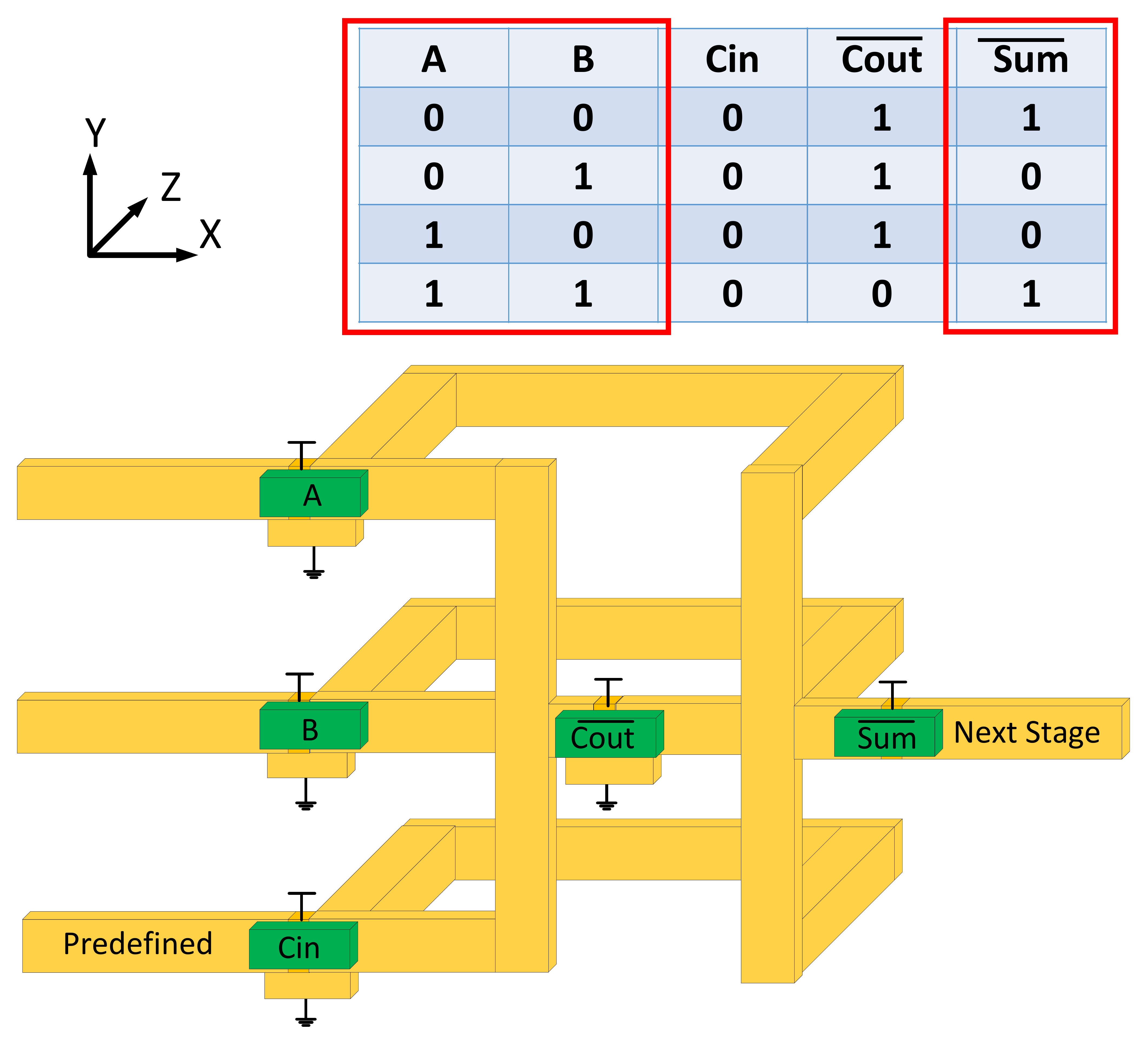}
\caption{1-bit full adder used as XNOR\cite{adder}. In the 2D implementation of this work, X and Y wires are in-plane metal wires and connections along the Z axis are vias.}
\end{figure}
The two inputs to this block (A and B) are coming from distinct sources. One of the inputs comes from the input image synchronized with the control voltage and the other input is given to the circuit during the learning phase. Compared to a CMOS counterpart, this structure exhibits very important advantages. First, it requires 5 magnets whereas the CMOS version requires at least 8 transistors for XNOR implementation. Second, this circuit has the capability of storing the training information without extra static power consumption, whereas in CMOS, excess power is consumed to store this data \cite{Fabrication}. Taking advantage of the non-volatile operation in ASL devices, the input magnets of this circuit can store the binary data and later the stored information is used to determine the magnetization direction of the next stages. Fig. 6 shows the simulated output waveform ($\bar{{sum}}$ magnet) of the XNOR block for different scenarios of input magnetization. As it is important to consider the breakdown current effects \cite{break}, we choose the 5mV supply voltage in our simulations. This is to ensure that the current density is safely below the breakdown value. It is noteworthy that for channels with higher breakdown current densities, higher voltages can be applied and the operation speed increases. In this simulation, a control voltage of 5mV is applied on the magnets at $t=0$. The total power consumption of the XNOR gate is 11$\mu W$ and the estimated area is less than 0.3$\mu m^2$. As we apply a control voltage on the XNOR gate, the output magnetization remains in $-X$ orientation (initial condition of magnetization in this simulation) if the pixel values are different. In case of having similar inputs to this gate, the output magnetization switches to $+X$ direction as shown in Fig. 6. We have to clarify that the initial condition of output magnet does not change the final magnetization orientation.
\begin{figure}[!h]
\centering
\includegraphics[width=.75\textwidth]{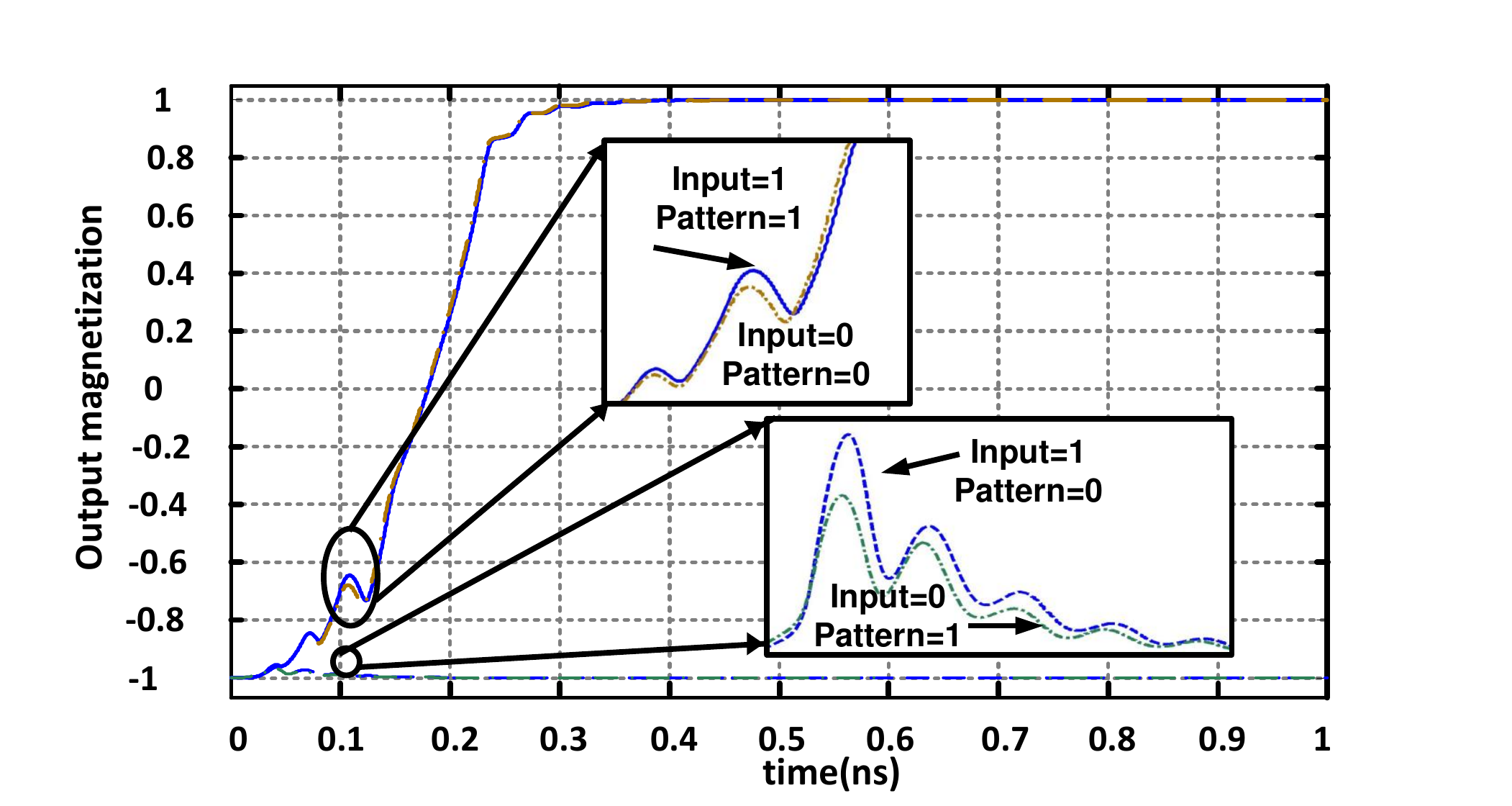}
\caption{Simulated output waveforms of XNOR gate}
\end{figure}
%After the output magnetization orientations of the majority gates settle to steady state values, the single pixel comparators operate. These gates determine the correlation of input pixel and corresponding mean pixel. The output magnet of each majority gate is connected to the subsequent XOR gate through a short interconnection. The reason we use short interconnections in this architecture is to make sure that the delay is minimized. This stage is consisted of all of the XOR comparator blocks implemented from \cite{adder}. A total number of \textit{P} XOR gates (\textit{P} determines the number of pattern images) are used for each pixel and the input data will be applied synchronously to all the \textit{P} gates, after the preset signal of the input magnet is enabled. The other input of XOR gate comes from the pattern data which has been stored in the system memory. In this transition, the output decision of each XOR gate has to identify the matching between the input pixel and the corresponding pattern pixel.
\subsection{Construction of the mean pixel}
As a reliable and simple way to extract the information from the training set, we construct the mean image as discussed in Section III. The ASL majority gate with the schematic shown in Fig. 1(b), provides a low power and efficient implementation of the mean image. The inputs to this majority gate, come from $P$ different users and the images that system receives during the learning phase are constrained to be mainly similar along the rows. By applying the control voltage on the magnets of this gate, the output magnetization either switches to other value or remains in the same magnetization orientation. If the applied control voltage is negative, the output final magnetization orientation is the majority of the input magnetizations. In the case of positive control voltage, the output magnetization settles to the complementary majority value of the input magnetizations. For this system, since we apply unified positive voltages, the majority gates settle to the complementary majority value. In order to extract more information from the majority gates operation in this circuit, we assume a unified value of initial magnetization orientation on the output magnets of each stage of majority gates. This enables us to recognize the total count of matches or mismatches between the input magnetizations to each majority gate, as we will discuss later. The total power consumption of each majority gate in this circuit is 3.75$\mu W$ and the estimated area is less than 0.2 $\mu m^2$.\\

\subsection{Single Pixel Comparator}
By having the required blocks, we propose the two different versions of the single pixel comparator. 
\subsubsection{Standard implementation}
The schematic of this implementation and the table with the detailed operation are shown in Fig. 7. This circuit operates in the same order discussed in Section III. The first stage of the circuit is a majority gate with inputs coming from the $P$ users in the learning phase. The output of this majority gate settles to the corresponding mean pixel value. The output of this gate is connected to a comparator circuit which has the other input coming from the input image. The connection is through a short metallic interconnect to minimize the delay. When the learning phase is over and the detection phase starts, by applying the control voltage across the magnets of the comparator circuit, the ``Pixel" magnet settles to the comparison value of the mean pixel and the input pixel. It is noteworthy that the input pixel can be applied on magnet $Q_{ij}$ after the $\bar{P_{ij}}$ magnetization settles to the mean pixel; hence, no extra memory circuit is required to store the value of $\bar{P_{ij}}$. 
\begin{figure}[!h]
\centering
\includegraphics[width=.7\textwidth]{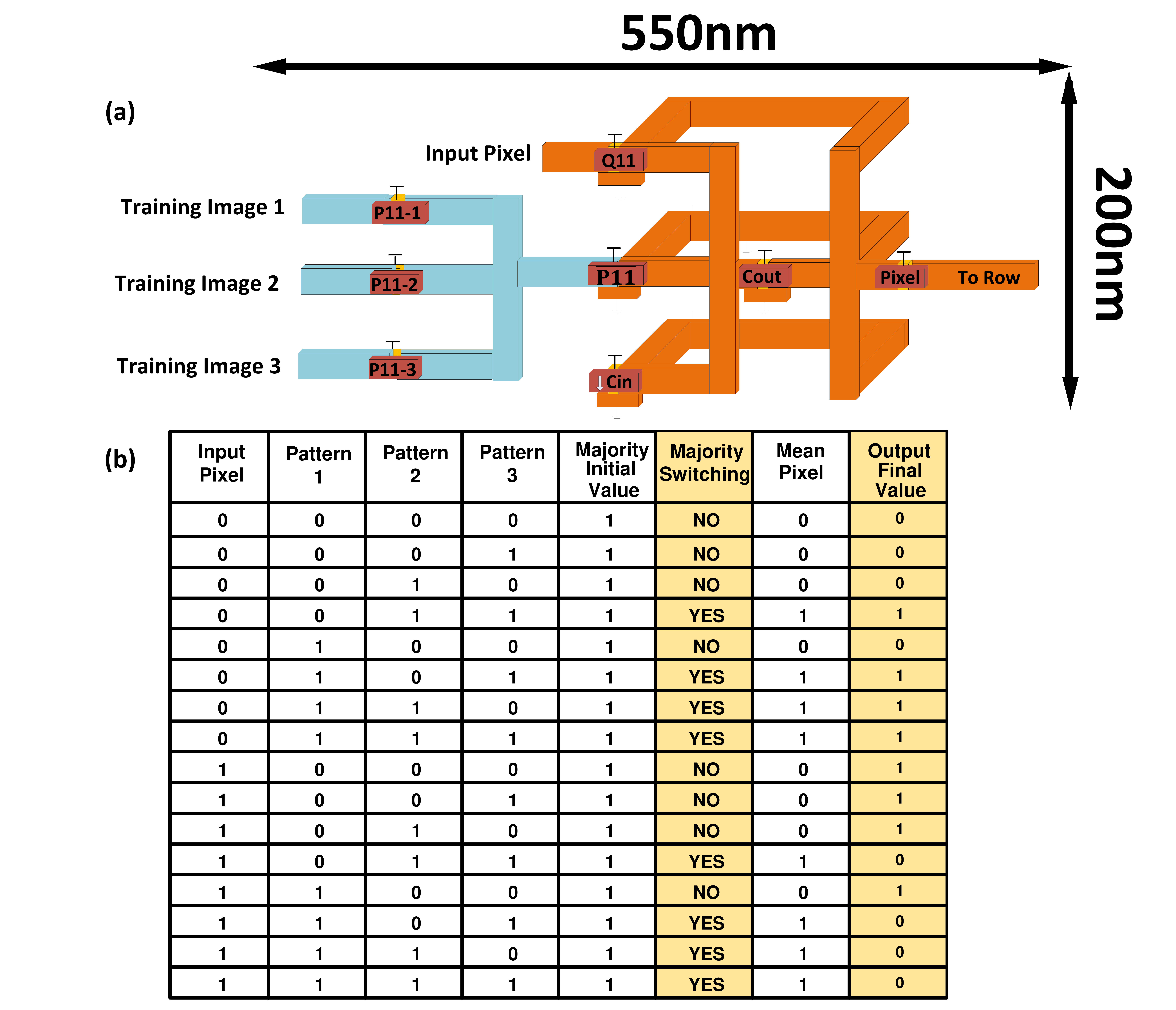}
\caption{(a) Standard single pixel detector schematic. (b) The truth table with the detailed operation of the circuit.}
\end{figure}
\subsubsection{Comparator-First implementation}
In this version, there will be the same number of comparator circuits as the total number of training images at the input side. The comparators have the input image pixel, $Q_{ij}$ in common and differ in their other input which comes from their corresponding training image. The output magnets of the comparators are connected to the Pixel magnet through metallic interconnects in a majority gate configuration. During the learning phase, the pattern pixels are stored in the corresponding input magnets. By applying the control voltage on the magnets of the circuit, the detection phase starts and the ``Pixel" magnetization settles to the comparison value of the mean pixel and the input pixel. The schematic of the circuit and the detailed operation table are shown in Fig. 8.\\
\begin{figure}[!h]
\centering
\includegraphics[width=.65\textwidth]{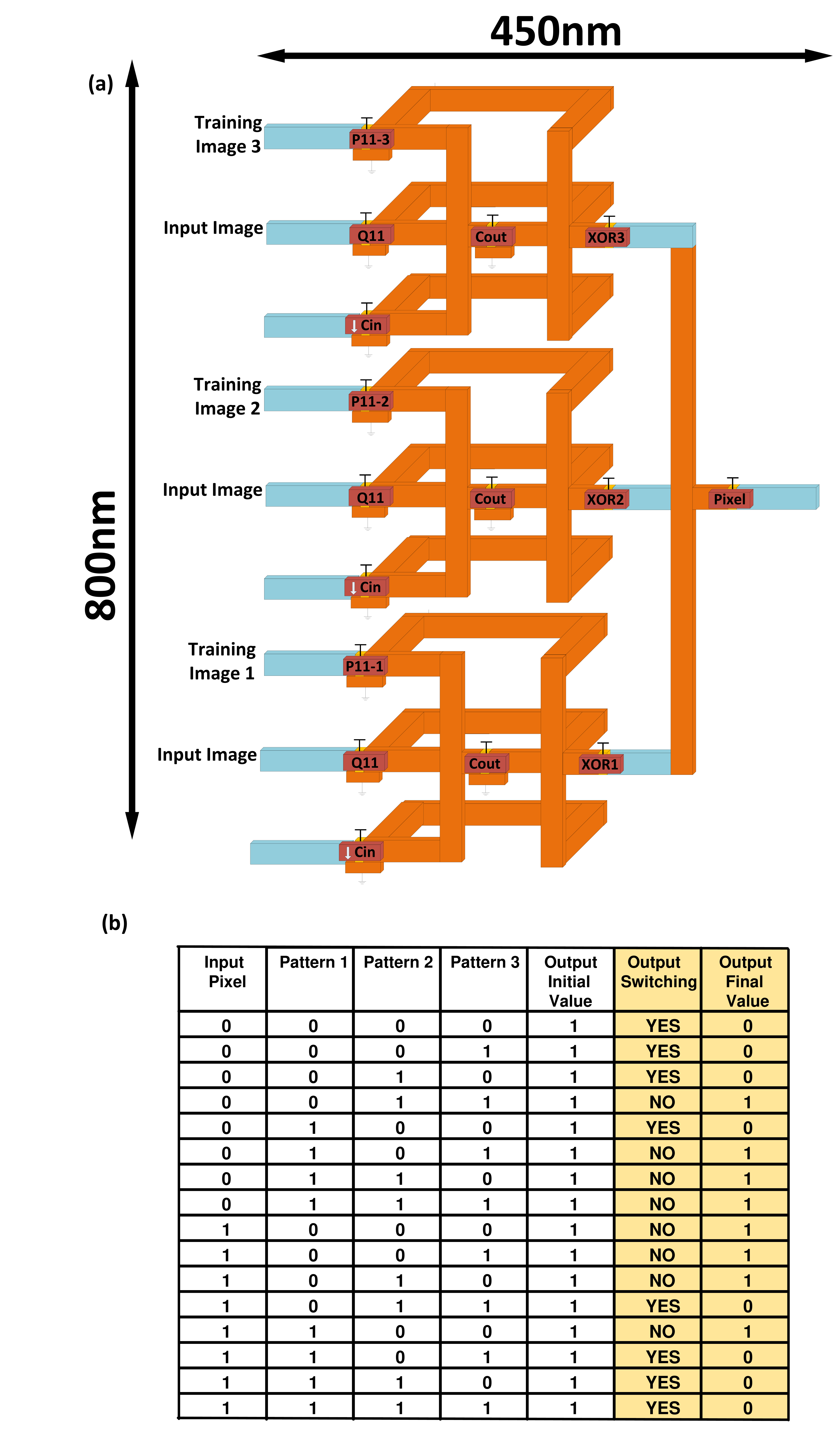}
\caption{(a) Comparator-first pixel detector schematic. (b) The truth table with the detailed operation of the circuit.}
\end{figure}
As it can be verified by comparing the last columns of Fig. 7(b) and Fig. 8(b), the ``Pixel" steady state value is identical in the two versions. To verify the identical output result from the two different versions of the implementation in a more general case, we have to prove that the majority operation and the comparison (XOR/XNOR) operation are interchangeable, i.e.,\\\\
\textbf{Proposition 1}\textit{ Given $x$, $y_1, y_2,\cdots, y_P$ as binary variables and $P$ as an odd integer number,}
\begin{equation}
x\oplus nint(\frac{1}{P}\sum\limits_{k=1}^P y_k)= nint(\frac{1}{P}\sum\limits_{k=1}^P (x\oplus y_k)),
\end{equation}
where $\oplus$ denotes the XOR operation. The mathematical proof of this proposition is shown in Appendix A.\\

Although the standard implementation has slightly lower power consumption (Less number of devices) and a smaller area, we select the comparator-first design as the unit cell of this circuit. This is due to the fact that the output magnetization transient of this circuit provides more information on the similarity of the training pixels and the input pixel. Based on Fig 7(b) and Fig 8(b), the final value of output magnetizations, in the two cases are identical. However, the Comparator-first output magnetizations is coming from a majority gate and switches when the majority of pattern pixels have the same value of the input pixel.  If the majority gate at the output of Comparator-first circuit has a low fan-in ( e.g., $\leq 5$), the switching transient behavior will be less sensitive to the accumulated thermal noise and the information on the number of training pixels with identical values will be provided. On the other hand, in the standard implementation, the output magnetization is from the XNOR circuit and conveys no information on the number of similar pattern pixels. Based on Fig 7(b), the output magnetization transient will not add information on the number of training pixels with identical values. This is particularly important when the user in the detection phase tracks the total count of pattern pixels with the similar value.

\subsection{Non-Boolean Row Decision-Maker}
The last stage of the proposed circuit uses the interesting feature of the ASL majority gate as a means to quickly decide about the mainly similarity of the input image and the mean image, along the rows. The inputs to this majority gate is from the ``Pixel" magnets of the pixels along the same row of the image. The connection is through short interconnects to minimize the delay.  As mentioned before, the spin torque transferred from the input magnet to the output magnet in the ASL majority gate, is determined by the magnetization of the input devices. As the number of devices with similar magnetization orientations increases, the transferred spin torque increases; hence, the output magnetization switching becomes faster according to (3). By proper selection of the control voltage timing and also the dimensions of the nanomagnets and metallic interconnects in this gate, a reliable decision-making based on the transient behavior of output magnetization is achieved. This final majority gate is sensitive to the uncorrelated thermal noise of input magnets; hence, an intentional low fan-in number ($\leq 5$) has to be selected. In our simulations, 3 magnets from the previous pixel stages are connected to this gate and as it will be shown in simulation results, a reliable decision-making is achieved. \\

The complete circuit for the full image comparison consists of two stages. The unit pixel comparator and the row majority gate. The structure consisting of the comparator-first circuits and the Row majority gate for a $3\times 3$ pixel image comparison is called the ``Smart Detector Cell''. This naming convention, helps the discussion of operation in the next section. We call these detector cells smart because they can perform multiple tasks of ``storage'', ``Boolean Computation'' and ``non-Boolean decision-making'' in a time-efficient manner. The schematic of this circuit is shown in Fig. 9. The total power consumption of this circuit is 115 $\mu W$ and the occupied area is less than 0.5 $\mu m^2$.

We have to mention that in our simulations, we have taken into account the effect of magnetization switching. The ASL device acts like a resistive network and the power consumption will not change with time. In these devices, the current passes through only one magnet and therefore does not change with time and the magnet switching of the input side, will influence the switching delay of the last magnet. In this paper, we consider the worst case delay which takes into account the switching delay from the input magnet of the first device to the output magnet of the last device as well as the transport delay within the metallic interconnects. For DC power consumption estimations, we have performed DC and transient simulations and the results are consistent. It is noteworthy that the low operational voltage of the circuit will lead to a low power consumption.

In a real implementation of this work, read/write circuits are added to fully realize the circuit. However, this paper focuses on the processing circuit without concerns regarding the feeding and extraction of the input and output data. In order to feed the input data, spin polarized currents are used to initialize the magnetization of input magnets based on the training images, similar to \cite{Sharad}. On the other hand, the number of write units is equal to the number of pixels, while there is one output, which translates to small overhead. The decision data is in form of time delay and can be stored on a capacitor, where the delay impacts the amount of the stored charge. The other possibility to extract he output data will be using MTJ devices, as mentioned in \cite{MTJ}.

\begin{figure}[!h]
\centering
\includegraphics[width=.6\textwidth]{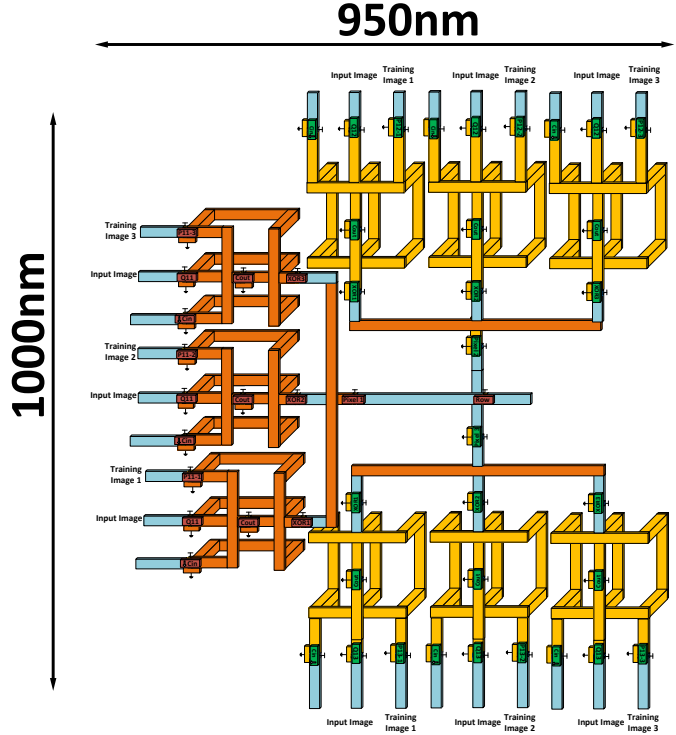}
\caption{Structure of the unit smart detector cell}
\end{figure}

%\subsection{Fan-in and Fan-out considerations}
%As mentioned before, the Fan-in and Fan-out numbers to different stages are determined by certain considerations. By increasing the number of inputs to a majority gate, the uncorrelated thermal noise of all the input magnets add up; hence, the transient behavior of output magnet becomes less predictable.  As shown in next section, in this proposed architecture, the first stage of majority gates receive signals from \textit{P} different users, \textit{P} representing the number of training images. Based on the user preferences, the transient of output magnetization in these majority gates will not be monitored if only the steady-state value of magnetization is of interest. This will be the case when the users, do not care about the number of similar training values for a single pixel. However, if the number of similar pixels has to be identified, the transient magnetization has to be immune enough to the effect of thermal noise. This will dictate a less fan-in number to these majority gates.\\

%In  the second stage of majority gates, the decision making for a row is performed based on the output magnetization transient. Due to this criterion, a low fan-in is considered for these majority gates.\\

\section{Simulation Results}
In this section, we provide two different examples to show the reliable performance of smart detector cells.
\subsection{Non-Boolean Hamming Distance Identifier of 3$\times$3 Pixel Pattern and Input Image}
In this example, we only have one training image and one input image. To compare the similarity of these two images, we need 9 XNOR gates to identify the similarity of corresponding pixels in the two images and 3 majority gates with Fan-in of 3 to decide on the similarity of the corresponding rows. It is also obvious that the mean image in this case will be the same pattern image. The smart detector cell in Fig. 9, has 3 comparator-first circuits and a \textit{Row} majority gate. The mainly similarity of the rows can be determined by the Pixel majority gates. The last majority gate in this case, settles to $+X$ magnetization if at least 2 rows are mainly similar. However, in this simulation, the output magnetization of this gate is not important. In this simulation, the initial magnetizations of the comparators and the majority gate outputs are set to 1. Fig. 10 shows the two images as well as the transient magnetization for various magnets. The Pixel waveforms overlap in some cases and that is why we are showing only 3 pixels in this figure. As expected, the comparator outputs switch for $P_{21}$ and $P_{22}$ pixels since the values in the input image and the pattern image are different. For the rest of pixels, the comparator output is $+X$ magnetization and will not switch. Subsequently, row 1 and row 3 both exhibit perfect similarity and the output of the corresponding majority gates switch within the shortest time as it can be compared with Fig. 2(b). On the other hand, row 2 exhibits a mismatch and therefore can not switch to $-X$ magnetization orientation. The control voltage of 5 mV is applied on all the magnets at $t=0$ and the circuit compares the two images in less than 0.6 ns. Compared to CMOS circuits, this exhibits a much lower operational voltage and decision time.

\begin{figure}[!h]
\centering
\includegraphics[width=.65\textwidth]{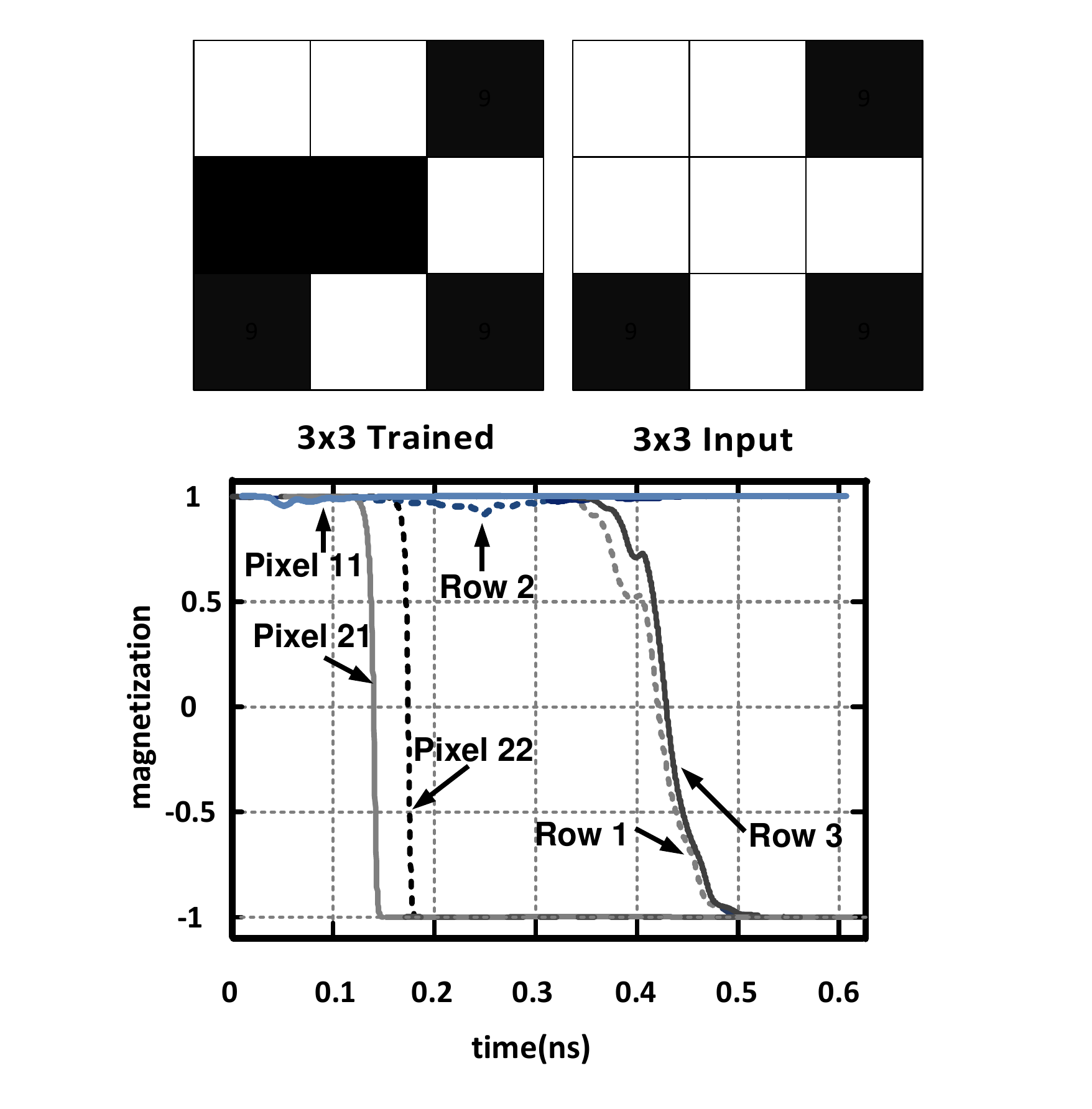}
\caption{Using a single smart detector cell, we can compare these $3\times 3$ pixel images. The waveforms of the comparators and majority gates (bottom).}
\end{figure} 
\subsection{Non-Boolean Similarity Comparison of a 9$\times$9 Pixel Image and a Set of 3 Pattern Images}
In order to incorporate the smart detector cells for larger images, we need an accurate design of the cells. Here, we develop a circuit for training with $9\times 9$ pixel images and perform a non-Boolean comparison between the constructed mean image and the $9\times 9$ pixel input image. In this simulation, 3 different users write the word ``Spin'' by their own choice of pixels. The 3 pattern images are shown in Fig. 11.\\
\begin{figure}[!h]
\centering
\includegraphics[width=.65\textwidth]{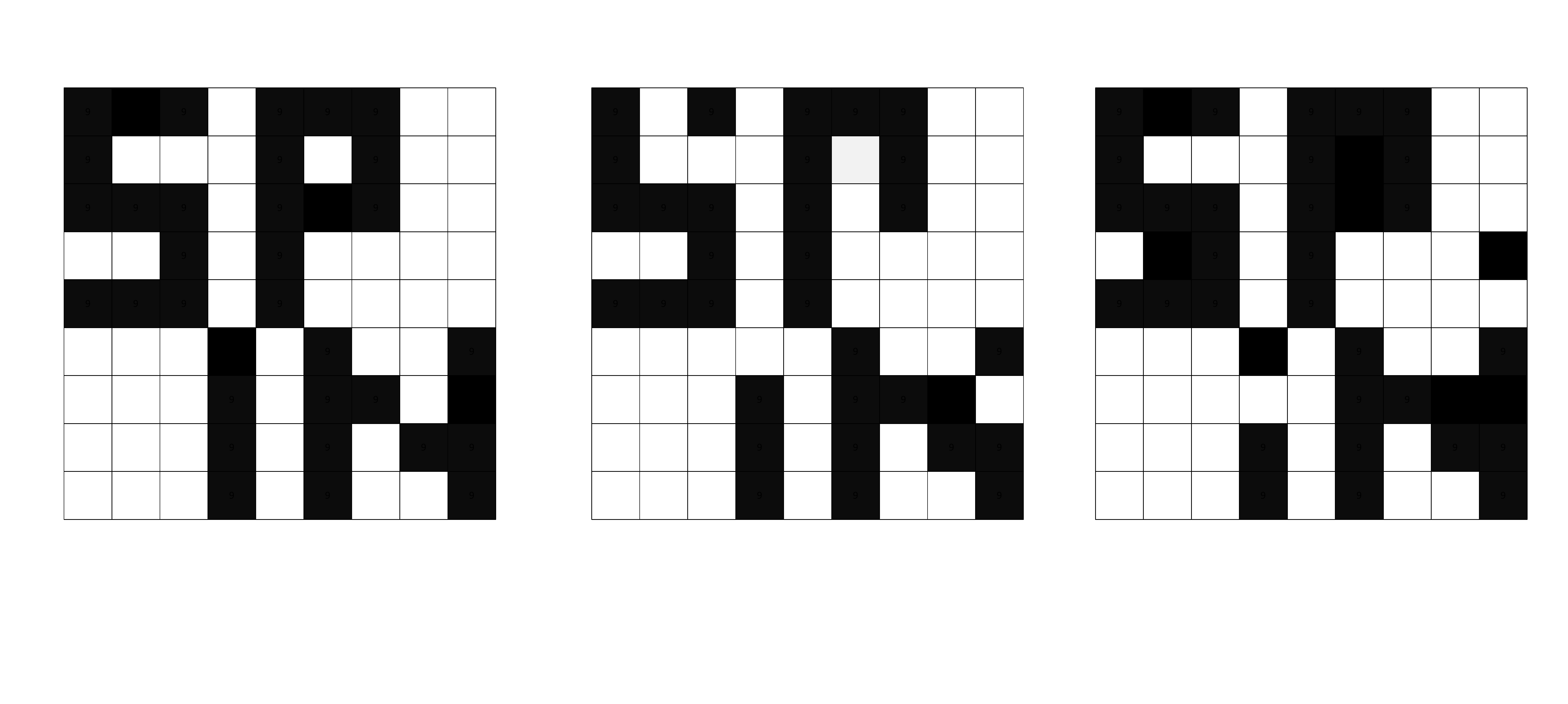}
\caption{Training set for the $9\times 9$ pixel images}
\end{figure}

In the detection phase, a new user of the circuit, chooses an arbitrary image of interest as the input. As an example in this simulation, the user chooses the word ``swim'' as shown in Fig. 12 (left). The circuit should compare this image and the mean image constructed from the training set.\\
 
\begin{figure}[!h]
\centering
\includegraphics[width=.75\textwidth]{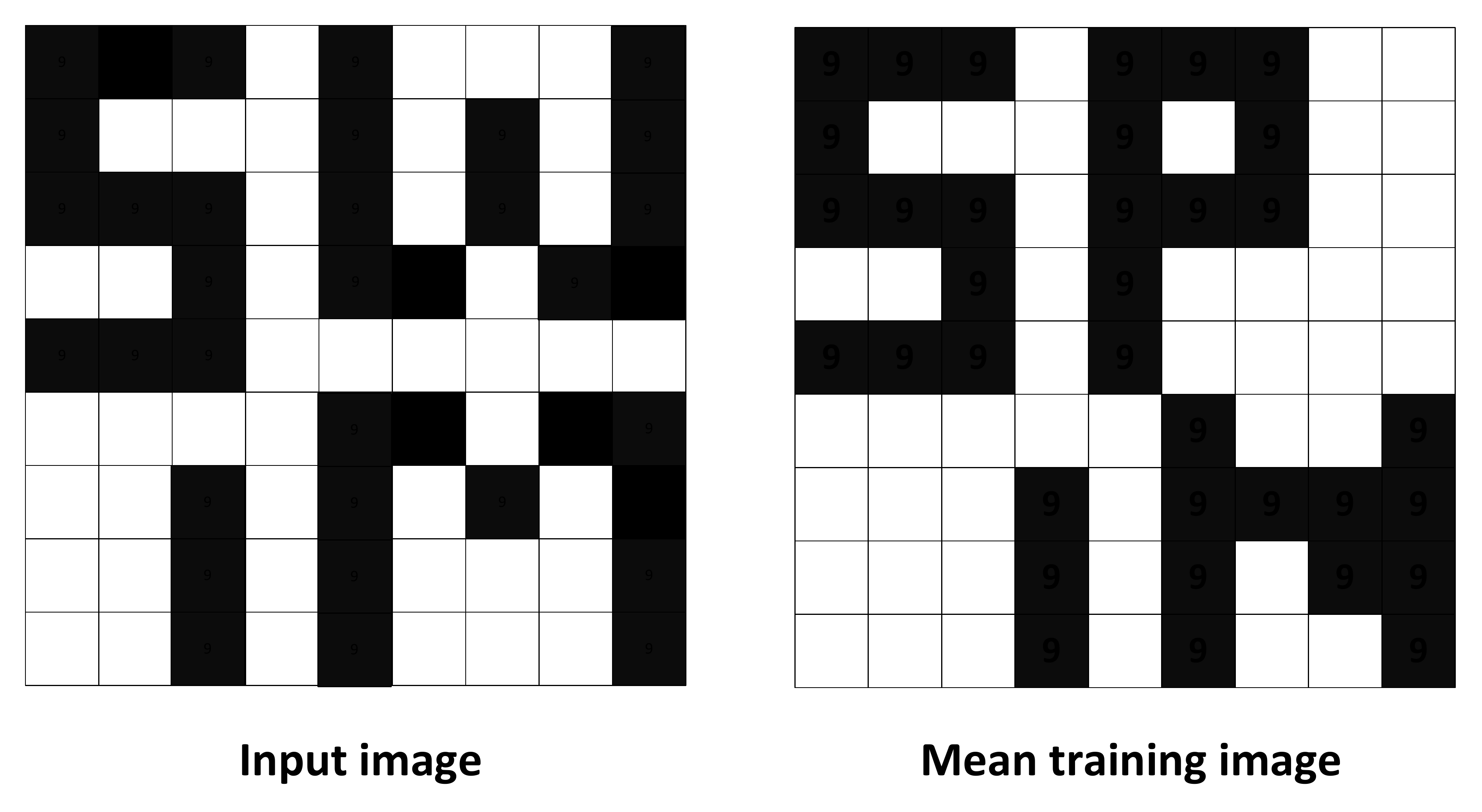}
\caption{The input image (left) and the mean image (right).}
\end{figure}
 The mean image of the training set is also shown in Fig. 12 (right). One particular advantage of constructing the mean image can be discussed here. As it can be seen in the mean image, those pixels which are mistakenly valued by a single user (e.g., $P_{26}$ and $P_{49}$) in the learning phase, are automatically corrected when the mean image is constructed. This is specifically useful, when the users during the learning phase, train the system with multiple versions of an image to make sure that the mean image represents their desired pattern. The mistaken values could be due to any source of error or distortion. In an ideal case where the thermal noise effect can be ignored, by simply changing the fan-in of different stages in the smart detector cell, the circuit can compare these two large images. However, in our simulations, as we model the thermal noise accurately, the fan-in considerations mentioned before, are particularly important. Based on these considerations, we break these 9$\times$9 images into smaller 3$\times$3 subimages, where a single smart detector cell unit can be used for the comparison. The 9 smart detector cells can operate in parallel and the circuit configuration can be determined by the user.  By this breakdown, we can also achieve more information on the pixels as we can check the mainly similarity for smaller blocks of the original image. The breakdowns of the mean image (squares on the right) and the input image (squares on the left) are shown in $3\times 3$ partitions in Fig. 13.\\
 
 \begin{figure}[!h]
\centering
\includegraphics[width=.65\textwidth]{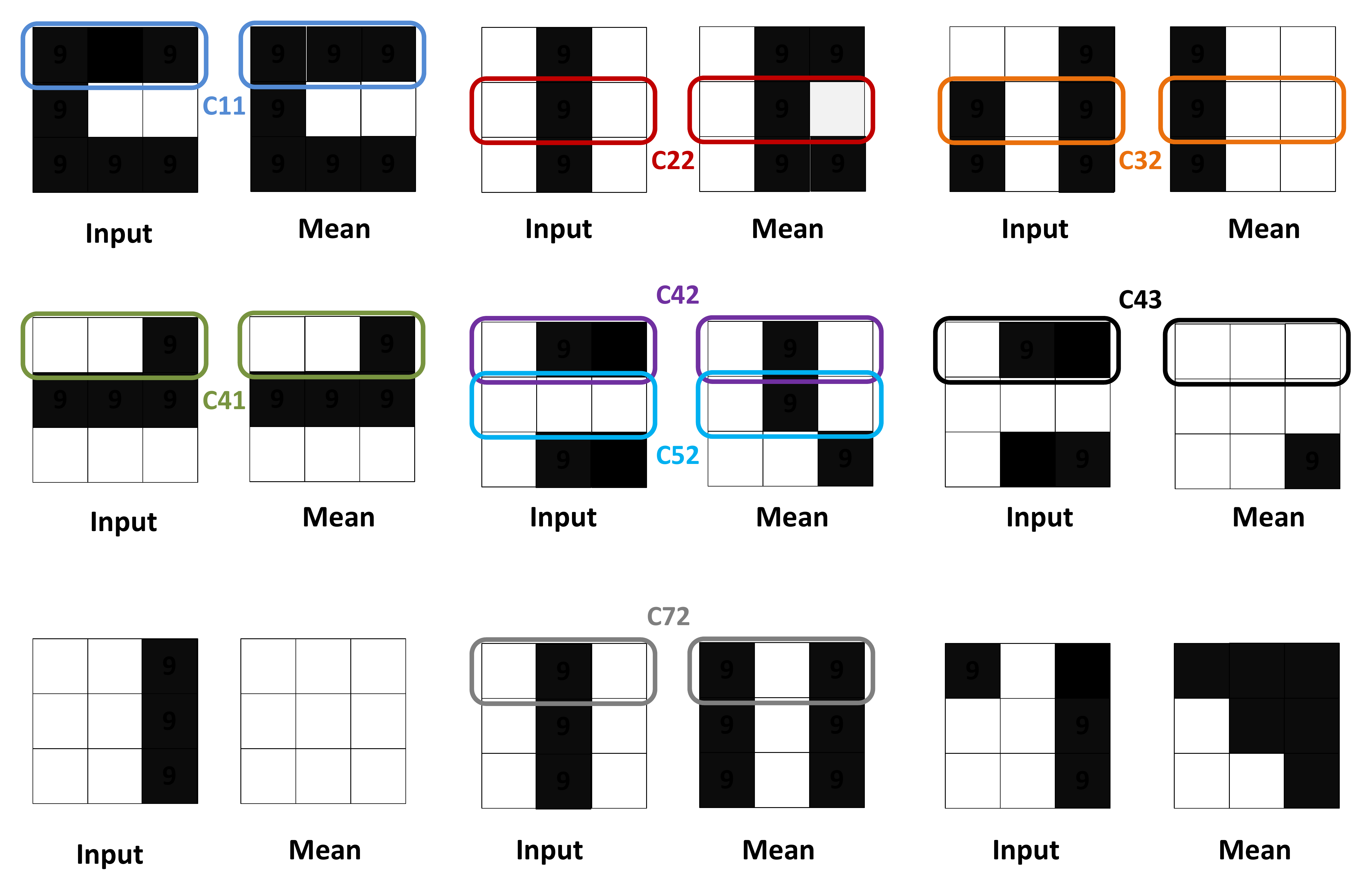}
\caption{Due to fan-in considerations, the circuit is consisted of 9 smart detector cells. The corresponding breakdowns of the mean image and the input image are shown here .}
\end{figure}
  In order to distinguish the different rows of smaller blocks, we use the notation of $C_{ij}$ clusters, which represents the elements of the $i^{th}$ row from column $3j-2$ to  column $3j$. The magnetization waveforms shown in Fig. 14 and Fig. 15 separately show the output magnetizations of smart detector cells for various clusters. The unified initial condition of the output magnet in this simulation is $-X$ magnetization orientation. In Fig. 14, the switching delay of output magnetizations for the clusters with perfect match ($C_{11}$, $C_{22}$ and $C_{41}$) and those with 1 mismatch ($C_{52}$, $C_{42}$ and $C_{32}$) can be easily distinguished. This phenomenon was previously described as the unique feature of ASL majority gates and helps the users to identify the number of mismatches along different rows. At the same time, the output magnetization of the clusters with the same level of similarity, are very close in time domain which makes this non-Boolean decision-making a reliable metric.
\begin{figure}[!h]
\centering
\includegraphics[width=.70\textwidth]{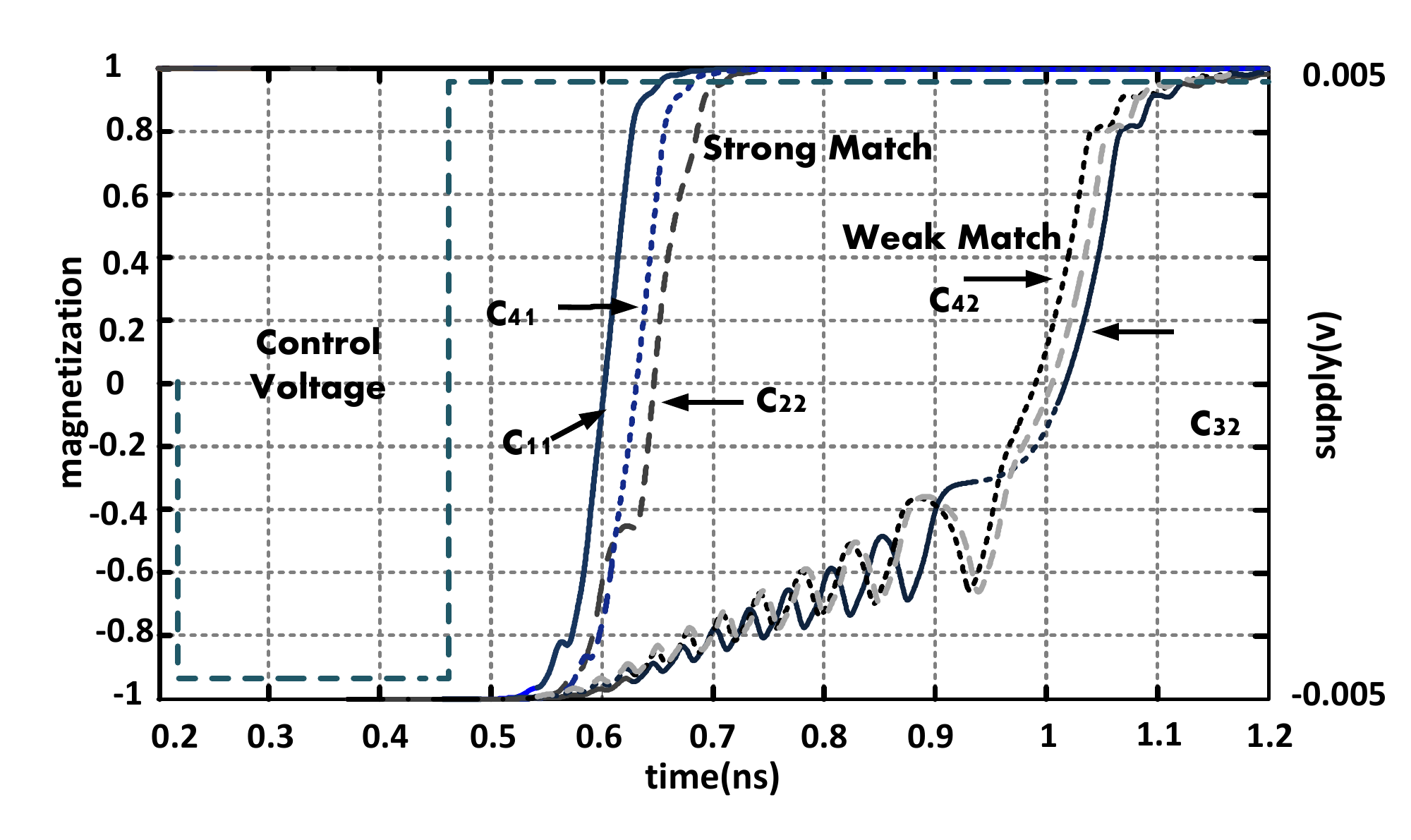}
\caption{The switching delay of output magnetization in last stage represents the similarity of input data and pattern data.}
\end{figure}  
On the other hand, in Fig. 15, the output magnetization can not switch for clusters with mismatches ($C_{43}$ and $C_{72}$) and as it can be seen, the level of precession for different mismatch levels is not the same. This is due to the different amount of spin torques provided in these two cases. If the user has a very high resolution study on the output magnetization, this can help to identify the number of mismatches; however, the switching transient is a more reliable metric and the same information can be extracted by repeating the simulation with the output magnet initial condition set to $+X$ magnetization.
\begin{figure}[!h]
\centering
\includegraphics[width=.70\textwidth]{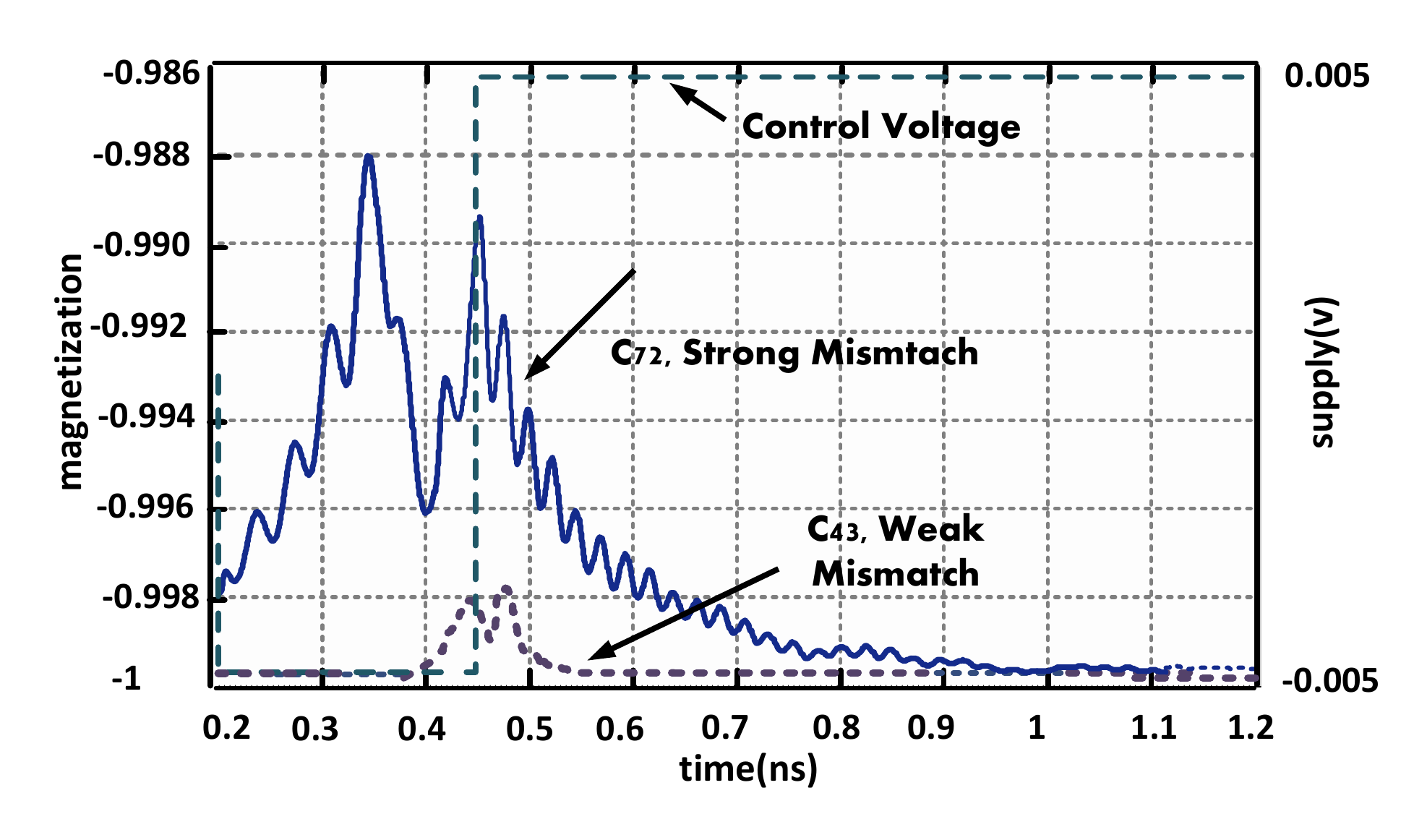}
\caption{Since these clusters represent mismatch, they can not switch and  the initial magnetization does not change. Note that the y-axis is showing from -1.002 to -0.998 in contrast with Fig. 14 in which the y axis is from -1 to 1.}
\end{figure}\\

As it can be seen in all the simulation results, this circuit can make a decision in almost 1000 ps for a $9\times 9$ pixel image, whereas in CMOS, this decision time, can not be less than few nanoseconds. For a detailed comparison between the two technologies, in table 1, the performance of this circuit and few existing CMOS circuits are compared.
\begin{table}[!t]
\renewcommand{\arraystretch}{1.3}
\caption{Performance Comparison with existing CMOS systems}
\label{table_example}
\centering
\begin{tabular}{|c|c|c|c|}
\hline
\bfseries Reference  &\bfseries \cite{table2} &\bfseries \cite{Sharad} &\bfseries This Work \\
\hline
Decision time  & 30ns &  N.A. & 1 ns \\
\hline
Image Size  & $32\times 32$ & 86 neurons &  $9\times 9$ \\
\hline
 DC Power  & N.A. & $2.2 mW$  &  990 uW \\
\hline
 Area  & N.A. & $0.018 mm^2$ &  $< 1 \mu m^2$ \\
\hline
Technology  & CMOS & Spin-CMOS & All-spin \\
\hline
\end{tabular}
\end{table}

\section{Conclusion}
We have presented a novel non-Boolean image recognition circuit based on all-spin logic devices. The introduced circuit can perform all the phases of a non-Boolean pattern recognition for binary images. Taking advantage of the non-volatility of ASL devices, the learning phase operation is performed incorporating no additional memory devices. By introducing the mainly similarity scheme, two different implementations of the circuit are proposed. As verified by simulation results, this circuit can recognize various sizes of binary image patterns faster than existing CMOS counterparts and consumes less power with an operational voltage of 5mV. Since the comparisons in this circuit are based on ASL majority gates, the computational complexity of the operation is also less compared to existing circuits. The proposed circuit has applications in fast and low power image recognition for security, medical imaging, and sensing. \\\\

% or


\section*{Appendix  A: Proof of proposition 1}
In this appendix, we mathematically verify (7). Since all the variables are binary-valued, 
\[0\leq \frac{1}{P}\sum\limits_{k=1}^P y_k \leq 1.
\]
The $nint$ operation results in 0, when
\begin{equation}
\sum\limits_{k=1}^P y_k < \frac{P}{2},
\end{equation} 
otherwise it results in 1.  Therefore, there are 4 different possibilities for the variables, as shown in table II. In order to simplify the notations, we also define, 
\begin{equation}
z_k=x\oplus y_k \qquad  \forall k\in\{1,\cdots, P\}.  
\end{equation}

\begin{table}[!h]
\renewcommand{\arraystretch}{1.3}
\caption{Possibilities of $x, y_1, \cdots, y_P$}
\label{table_example}
\centering
\begin{tabular}{c|c|c|c}

\bfseries $x$ &\bfseries $\sum\limits_{k=1}^P y_k$ &\bfseries $nint(\frac{1}{P}\sum\limits_{k=1}^P y_k)$ &\bfseries $x\oplus nint(\frac{1}{P}\sum\limits_{k=1}^P y_k)$  \\
\hline
0 & $< \frac{P}{2}$ & 0 & 0  \\
\hline
0 & $> \frac{P}{2}$ & 1 & 1  \\
\hline
1 &  $< \frac{P}{2}$ & 0 & 1   \\
\hline
1 &  $> \frac{P}{2}$ & 1 & 0  \\
\hline
\end{tabular}
\end{table}

Here, we verify (7) for the first row of table II. Using the same method for the other 3 rows, the proposition can be completely proved.\\

If $\sum\limits_{k=1}^P y_k < \frac{P}{2}$, fewer than $\frac{P}{2}$ of $y_k$'s are 1. Given $x=0$, this means that fewer than $\frac{P}{2}$ of $z_k$'s are 1 and the rest are zero, i.e.,
\[ \sum\limits_{k=1}^P z_k < \frac{P}{2}. \] 

Similar to (8), by applying the nearest integer function,
\[nint(\frac{1}{P}\sum\limits_{k=1}^P z_k)=0.\]
 
\section*{Appendix B: Simulation Parameters}
\begin{center}
\begin{tabular}{ |l|l|l| }
  \hline
   \multicolumn{3}{|c|}{ \textit{Size Effect Parameters}\cite{impact}} \\
  \hline
 Side Wall specularity& P & 0 \\
 Grain Boundary Reflectivity &R & 0.2  \\
 
 \hline
 \multicolumn{3}{|c|}{ \textit{Interface Parameters(Co/Cu)}\cite{Interface}} \\
  \hline
 Majority Spin conductance&$G_{\uparrow}$ & 0.375 1/$\Omega$ \\
 Minority Spin conductance&$G_{\downarrow}$ & 0.125 1/$\Omega$ \\
 Real Spin-Mixing Conductance&Re$G_{\uparrow\downarrow}$ & 3.43751/$\Omega$  \\
 Imaginary Spin-Mixing Conductance&Im$G_{\uparrow\downarrow}$ & 9.37$\times 10^{-3}$ 1/$\Omega$  \\
 \hline
  \multicolumn{3}{|c|}{ \textit{Ferromagnet(Co)}\cite{new}} \\
  \hline
 Ferromagnet Length&$L_x$ & 75.00 nm \\
 Ferromagnet Width&$L_y$ & 25.00 nm \\
 Ferromagnet Height&$L_z$ & 3.00 nm \\
 Gilbert Damping Coefficient&$\alpha$ & 0.0021 \\
 Gyromagnetic Ratio &$\gamma$ & 1.76$\times 10^{11}$ 1/sT \\
 Saturation Magnetization&$M_s$ & 1.45$\times 10^{6}$ A/m \\
 Number of spins in magnet&$N_s$ & 1.34$\times 10^{6}$ 1/V \\
 Energy Density&$K_u$ & 0.5$\times 10^5 J/m^2$ \\
 \hline
 
  \multicolumn{3}{|c|}{ \textit{Channel(Cu)}\cite{new_2}} \\
  \hline
Channel Length&$L_{int}$ & 212.5 nm \\
Channel Width&$W_{int}$ & 50 nm \\
Thickness/Width aspect ratio&AR & 2.0 \\
Channel Thickness&$H_{int}$ &100.0 nm \\ 
Cross section Area&$A$ &5000 $nm^2$ \\ 
Finite difference spacing &$\Delta x$ & 10.0 nm \\
Conductivity&$\sigma$ & 41.549 $1/\mu\Omega m^2$\\
Diffusion coefficient & D & 0.014 $m^2/s$\\
Permeability& $\mu$ & 0.003 $m^2/Vs$\\
Spin relaxation time & $\tau_s$ & 10.939 ps \\
 \hline
\end{tabular}\\
\end{center}
\section*{Acknowledgment}

The authors would like to thank Mr. Vahnood Pourahmad, Dr. Abolhassan Vaezi and Dr. Mohammad Emadi from Cornell University and Dr. Alireza Aghasi from Georgia Institute of technology for helpful discussions. We also, like to thank Semiconductor Research Corporation (SRC) and Institute for Nanoelectronics Discovery and Exploration (INDEX) for the financial support of this project.




%\bibliographystyle{IEEEtran}


%\vfill


\end{document}